\documentclass[review,number,sort&compress]{elsarticle}

\usepackage{lineno}

\linenumbers

\usepackage{amssymb,amsmath}    % need for subequations
\usepackage{graphicx}   % need for figures
\usepackage{color}
\usepackage{hyperref}   % use for hypertext links, including those to external documents and URLs

\journal{Nucl. Instrum. Meth.  A}

\begin{document}
    \begin{frontmatter}
        \title{Novel techniques to search for neutron radioactivity}
    %%%%%%%%%%%%%%%%%
    %% AUTHOR LIST %%
    %%%%%%%%%%%%%%%%%
    \author[nscl,msu]{M.~Thoennessen\corref{cor1}}
    \ead{thoennessen@nscl.msu.edu}
    \author[TRIUMF]{G.~Christian}
   \author[nscl,chem]{Z.~Kohley}
   \author[nscl]{T.~Baumann}
    \author[nscl,msu]{M.~Jones}
%    \author[nscl,msu]{Z.~Schaedig}
    \author[nscl,msu]{J.~K.~Smith}
    \author[nscl,msu]{J.~Snyder}
    \author[nscl,msu]{A.~Spyrou}

    \cortext[cor1]{Corresponding Author.}
    %%%%%%%%%%%%%%%%%%
    %% ADDRESS LIST %%
    %%%%%%%%%%%%%%%%%%
    \address[nscl]{National Superconducting Cyclotron Laboratory, Michigan State University, East Lansing, Michigan 48824}
    \address[msu]{Department of Physics \& Astronomy, Michigan State University, East Lansing, Michigan 48824}
    \address[TRIUMF]{TRIUMF, 4004 Wesbrook Mall, Vancouver, British Columbia V6T 2A3, Canada}
    \address[chem]{Department of Chemistry, Michigan State University, East Lansing, Michigan 48824}  

    \begin{abstract}

Two new methods to observe neutron radioactivity are presented. Both methods rely on the production and decay of the parent nucleus in flight. The relative velocity measured between the neutron and the fragment is sensitive to half-lives between $\sim$1 and $\sim$100~ps for the Decay in Target (DiT) method. The transverse position measurement of the neutron in the Decay in a Magnetic Field (DiMF) method is sensitive to half-lives between 10~ps and 1~ns.

    \end{abstract}


    \end{frontmatter}

    \section{Introduction}
Nuclei with extreme neutron deficiency or neutron excess can decay by the emission of one or more protons or neutrons, respectively. The presence of the Coulomb barrier can significantly hinder the emission of a proton which can lead to fairly long lifetimes for this decay mode. The current status of one- and two-proton radioactivity has recently been reviewed by Pf\"utzner \cite{Pfu13}. In contrast, neutron emission typically proceeds on very short time scales ($\sim$10$^{-21}$~s), primarily due to the absence of the Coulomb barrier. However, it has been postulated that in special cases one- or two-neutron radioactivity might occur due to the presence of an angular momentum barrier \cite{Kry96,Gri11,Bau12}.
Lifetimes as short as 10$^{-12}$~s can be considered as radioactivity  \cite{Tho04} and only in the most extreme cases neutron emission is expected to reach these time scales. Thus traditional methods where the decaying nucleus is implanted in a detector and the subsequent decay is recorded are not applicable. With these methods the shortest measured lifetimes at present are 620~ns \cite{Lid06,Grz07} and 1.9~$\mu$s \cite{Bin05,Grz05,Ryk05} for $\alpha$- and proton-decay, respectively. 

Mukha {\it et al.} developed a new technique to measure the two-proton decay of $^{19}$Mg in flight by tracking the paths of the decay products and measured a half-life of 4.0(15)~ps \cite{Muk07}. Voss {\it et al.} studied the same decay with an adaptation of the recoil distance method. The degrader foil of a plunger device was replaced by a double-sided silicon strip detector to measure the energy-loss of the decaying nucleus ($^{19}$Mg) and the resulting fragment ($^{17}$Ne). The lifetime was then extracted from the intensity ratio of the energy loss peaks as a function of target to detector distance \cite{Vos11}.

In order to search for the corresponding decays of neutron-rich nuclei these or similar techniques have to be developed for neutron emission. Grigorenko {\it et al.} suggested one might adapt the tracking method by measuring the angular distributions of the neutron(s) and the fragment with high precision \cite{Gri11}. The reconstructed opening angle of the decay is then directly related to the decay energy. Caesar {\it et al.} extracted an upper limit of 5.7~ns for the decay of $^{26}$O by two-neutron emission by assuming the survival of $^{26}$O along the flight path in a magnetic field \cite{Cae12}. For future experiments they also proposed placing the target directly in front of a deflecting magnet and then deducing the lifetime from the horizontal position distribution of the neutrons. Most recently Kohley {\it et al.} applied another modification of the recoil distance method to analyze the previously reported ground-state two-neutron decay of $^{26}$O \cite{Lun12}. In this method the velocity difference between the neutrons and the fragments is sensitive to the lifetime if the decay occurs within the target. A half-life of 4.5$^{+1.1}_{-1.5}$(stat)$\pm$3(syst)~ps was determined for the decay of $^{26}$O \cite{Koh13}.

In the present paper we discuss calculations to extract the half-life sensitivities of the two methods mentioned above: the velocity difference method for the Decay in the Target (DiT) and the method to measure the horizontal neutron position distribution following the Decay in a Magnetic Field (DiMF).

\section{Test case of $^{16}$B}

In order to explore the feasibility of the DiT and DiMF methods for the measurement of neutron radioactivity, the single neutron decay of $^{16}$B was used as a test case. Although it is unlikely that neutron radioactivity will be observed in $^{16}$B, the reaction $^{17}$C(-p)$^{16}$B$\rightarrow^{15}$B + n was simulated because the complications due to the emission of multiple neutrons, occurring in the potentially more interesting case of the two-neutron emitter $^{26}$O, are avoided. 

$^{16}$B is unbound with respect to $^{15}$B plus a neutron \cite{Bow74,Lan85} and the ground state resonance was reported at 40(60)~keV \cite{Boh95} and 40(40)~keV \cite{Kal00} from multiparticle transfer reactions. Such a low decay energy (which is consistent with 0~keV within the given uncertainty) is critical for the possible observation of a finite lifetime because of the small barrier which is only due to the angular momentum. For an $l$=2 transition, which is expected for the ground state of $^{16}$B, a decay energy of  about 1~keV corresponds to a half-life of approximately 0.1~ps \cite{Kry96,Bau12}. Subsequently, invariant mass measurements found a resonance at low decay energies of 85(15)~keV \cite{Lec09} and 60(20)~keV \cite{Spy10}. This resonance most likely corresponds to the ground state decay. However, it is still conceivable that it could be a transition to one of the bound excited states of $^{15}$B. In an attempt to search for neutron radioactivity of $^{16}$B an upper limit for the half-life of 132~ps (68\% confidence level) was extracted from a proton stripping reaction from a radioactive $^{17}$C beam \cite{Kry96}.

\section{Decay in Target (DiT)}

    \begin{figure}
        \begin{center}
                \includegraphics[width=0.85\textwidth]{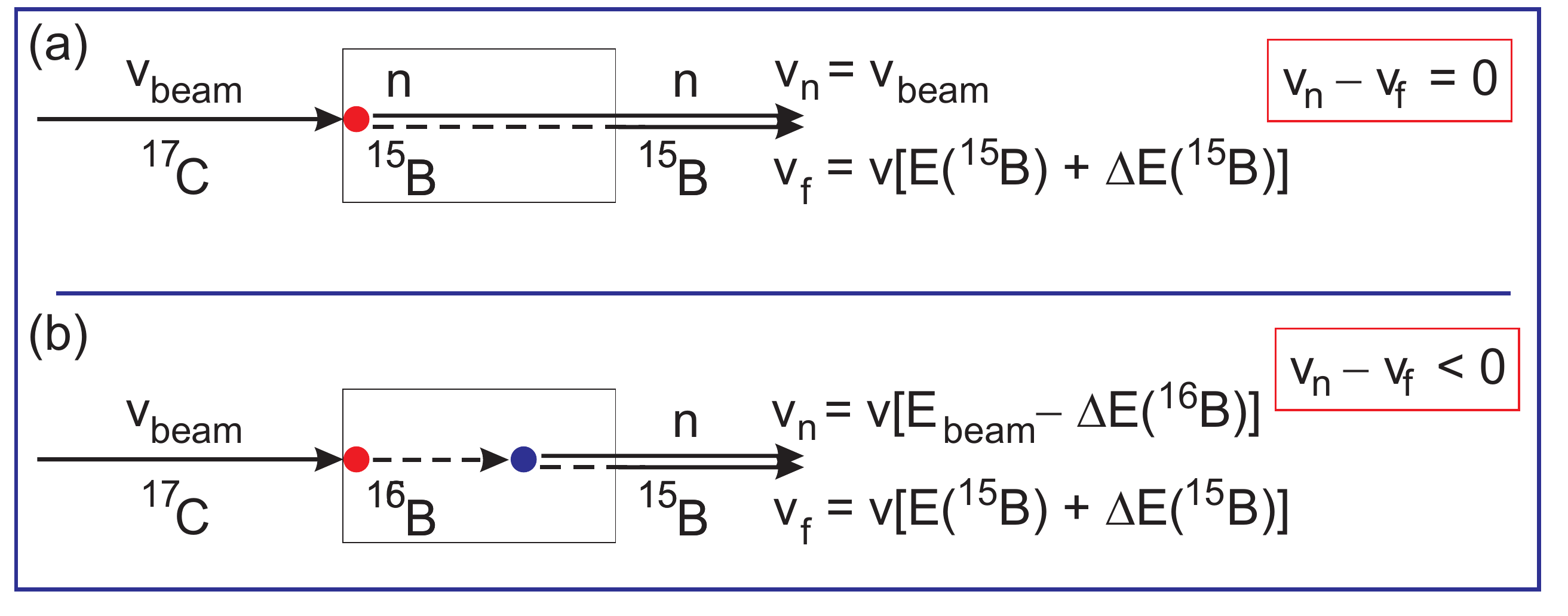}
        \end{center}
        \caption{Schematics of the DiT technique. Panel (a) shows the production of $^{16}$B and decay to $^{15}$B and a neutron at the beginning of the target, while in panel (b) $^{16}$B is produced at the beginning of the target but decays to $^{15}$B and a neutron at a later time in the target. Because the position of the decay is unknown, $\Delta$E($^{15}$B) is calculated assuming it traverses the whole target. The calculations of the relevant velocities are explained in the text.}
        \label{DiT:scheme}
    \end{figure}

The DiT technique was recently applied for the first time in the decay of $^{26}$O into $^{24}$O and two neutrons \cite{Koh13}. A schematic overview of this technique is shown in figure \ref{DiT:scheme} for the decay of $^{16}$B as an example. In this figure it is assumed that the one-proton removal reaction $^{17}$C(-p)$^{16}$B occurs at the beginning of the target and the outgoing fragment continues with essentially the same velocity as the incoming beam. As mentioned before, in order for long lifetimes to occur the decay energy for the subsequent decay of $^{16}$B into $^{15}$B and a neutron has to be very low and therefore recoil effects can be neglected. Panel (a) of the figure depicts the situation where the decay of $^{16}$B into $^{15}$B and a neutron proceeds instantaneously. While the neutrons continue at beam velocity through the target and into the neutron detectors, the $^{15}$B fragments lose energy as they traverse the target before the final energy is measured with charged-particle detectors. The fragment velocity at the interaction point can then be reconstructed from the measured final energy  [E($^{15}$B)] and the energy loss through the target [$\Delta$E($^{15}$B)] which can be calculated with the above assumptions from the incoming beam energy and the target thickness. The velocity difference v$_n -$ v$_f$ for this case is then equal to zero. If the decay occurs at a later time, when the $^{16}$B fragment has traveled through a fraction of the target, the calculated velocity difference will be less than zero as shown in figure \ref{DiT:scheme}(b). As the location of the decay is unknown, the fragment velocity is calculated with the assumption that the decay occurred at the beginning of the target, the same as in panel (a). The neutron velocity, however, will be reduced due to the energy loss of the $^{16}$B in the target before its decay to $^{15}$B and the neutron. The signature for a finite lifetime of the decay is thus a shift towards negative values of the velocity difference v$_n -$ v$_f$.

%    \begin{figure}
%       \begin{center}
%              \includegraphics[width=0.85\textwidth]{DiT-distributions.pdf}
%        \end{center}
%        \caption{Relative velocity (v$_n -$ v$_f$) distributions for the DiT technique assuming a $^{17}$C beam with an energy of 80 MeV/u. (a): The production of $^{16}$B is assumed to occur at the beginning of a 700~mg/cm$^2$ $^9$Be target for three different half-lives for the decay to $^{15}$B and a neutron. (b): The production of $^{16}$B can occur anywhere equally distributed throughout the target with an instantaneous decay to $^{15}$B and a neutron. The energy loss of the fragments through half the target is added back to fragments in order to center the relative velocity distributions. (c): Same half-live distributions from (a) assuming that the production can occur anywhere in the target. (d): Same distributions as (c) folded with the resolution of the velocity measurements.}
%        \label{DiT:dist}
%    \end{figure}

    \begin{figure}
       \begin{center}
              \includegraphics[width=0.6\textwidth]{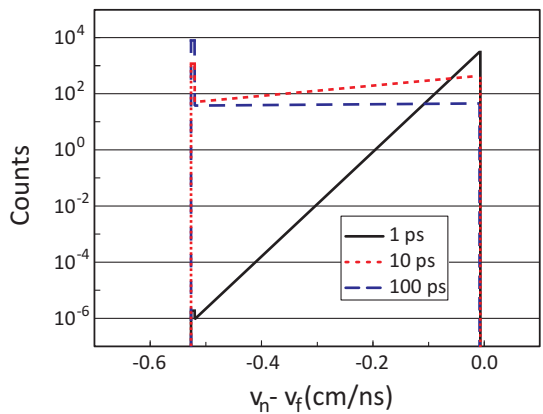}
        \end{center}
        \caption{Velocity difference distributions for the decay of $^{16}$B into $^{15}$B and a neutron at 80 MeV/u for three different half lives. The production of $^{16}$B is assumed to occur at the beginning of a 700~mg/cm$^2$ $^9$Be target. The fragment velocity was calculated after adding the energy loss of the fragment through the whole target to the final measured energy.}
        \label{DiT:fig1}
    \end{figure}

    \begin{figure}
       \begin{center}
              \includegraphics[width=0.6\textwidth]{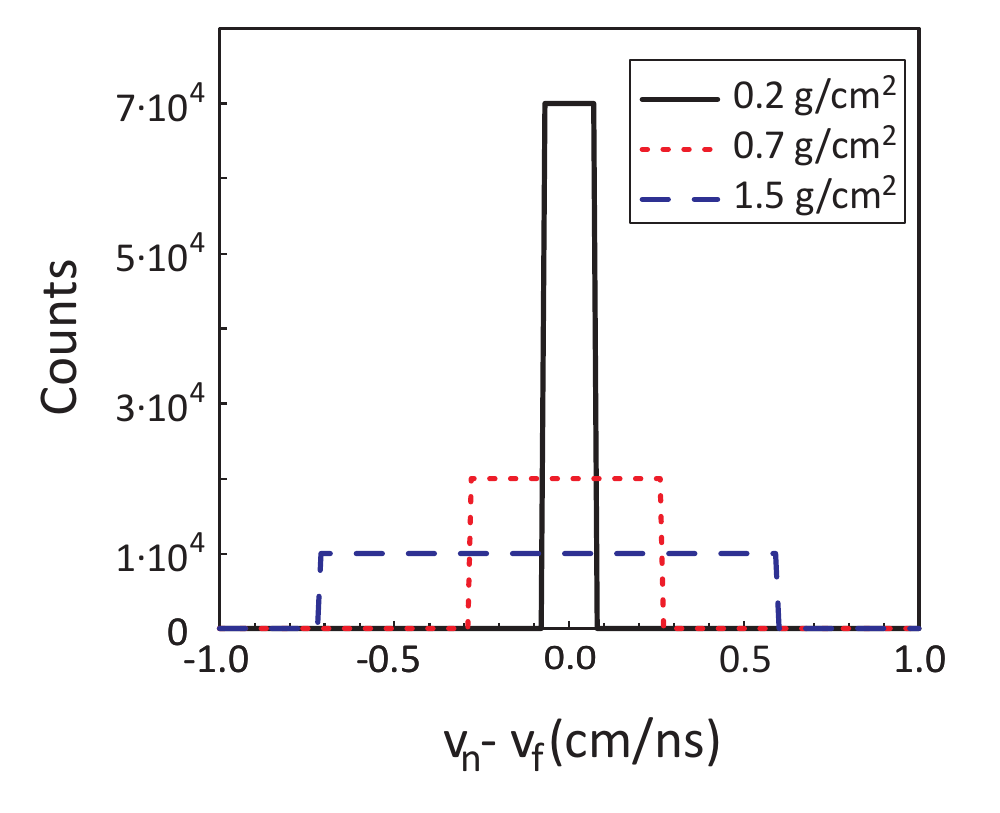}
        \end{center}
        \caption{Velocity difference distributions from the instantaneous decay of $^{16}$B (t$_{1/2}$ = 0~ps) to $^{15}$B for three different target thicknesses at an incident beam energy of 80 MeV/u. For these curves it was assumed that $^{16}$B could be produced (and decay instantaneously) anywhere equally distributed throughout the target. The energy loss of the fragments through only half the target was added back to the final measured fragment energy in order to center the velocity difference distributions at v$_n -$ v$_f \sim 0$.}
        \label{DiT:fig2}
    \end{figure}

    \begin{figure}
       \begin{center}
              \includegraphics[width=0.6\textwidth]{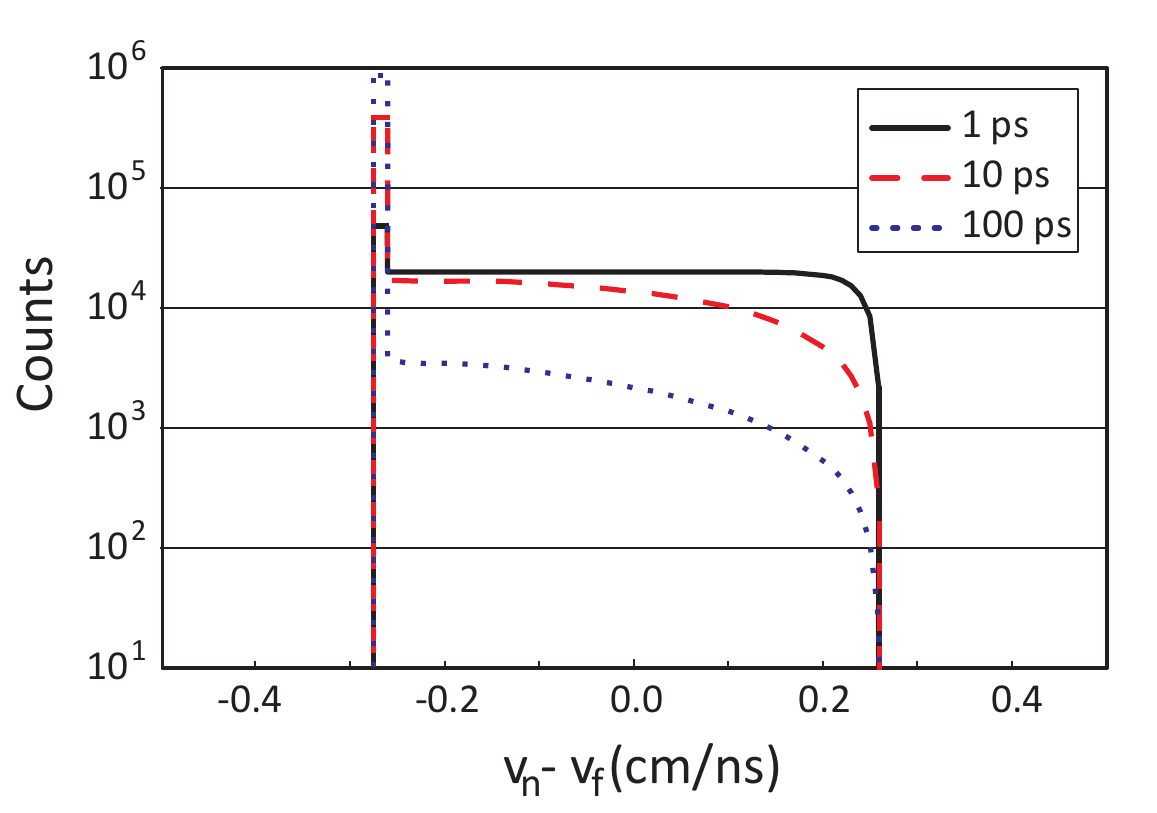}
        \end{center}
        \caption{Same as figure \ref{DiT:fig1} assuming that the production can occur anywhere in the target, i.e. the curves from figure \ref{DiT:fig1} were folded with the spread to the energy loss in the target shown by the red/dotted line (700~mg/cm$^2$) from figure \ref{DiT:fig2}.}
        \label{DiT:fig3}
    \end{figure}

\begin{figure}
       \begin{center}
              \includegraphics[width=0.6\textwidth]{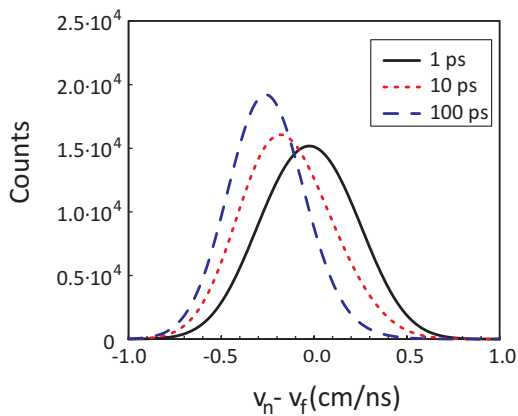}
        \end{center}
        \caption{Same as figure \ref{DiT:fig3} folded with the resolution of the velocity measurements.}
        \label{DiT:fig4}
    \end{figure}

The various effects contributing to the measured distributions are demonstrated in figures \ref{DiT:fig1}-\ref{DiT:fig4}. Figure \ref{DiT:fig1} shows the calculated velocity difference assuming that $^{16}$B is produced at the beginning of the target. Distributions for half-lives of 1~ps (black/solid line), 10~ps (red/dotted line), and 100~ps (blue/dashed line) are shown for a 700~mg/cm$^2$ thick $^9$Be target and an incoming $^{17}$C beam energy of 80~MeV/u. For the calculation of the velocity difference the energy loss through the full target was added back to the final fragment energy. The exponential decrease of the velocity difference translates directly to the half life. The beam energy corresponds to a velocity of 11.7~cm/ns and with a target thickness of 0.38~cm, the traversal time through the target is $\sim$32~ps. The decay rate for a half-life of 10~ps drops by about an order of magnitude during this time, consistent with the decrease shown by the red/dotted line in the figure. The sharp increase at the left edge of the distributions is due to the integral of decays outside of the target which is larger for the longer half-lives.

More realistic calculations have to take into account that the reaction can take place anywhere in the target. Because the exact interaction point is unknown, the velocity difference broadens due to the varying energy losses by the fragments in the target. This effect is shown in figure \ref{DiT:fig2} for target thicknesses of 200~mg/cm$^2$ (black/solid line), 700~mg/cm$^2$ (red/dotted line), and 1500~mg/cm$^2$ (blue/dashed line). The beam energy was 80~MeV/u and it was assumed that the subsequent decay occurs at the same time/place as the reaction (t$_{1/2}$ = 0~ps). In order to center the distributions at v$_n -$ v$_f$ = 0, only the energy loss through half the target thickness was added back to the final fragment energy.

The combined effect of finite half-lives and uniform distribution of the reaction within the target is shown in figure \ref{DiT:fig3} for the same three half-lives and conditions as in figure \ref{DiT:fig1}. Finally, the results of the simulations have to be folded with realistic resolutions of the detectors. The corresponding curves displayed in figure \ref{DiT:fig4} were calculated with an overall resolution of 0.2~cm/ns which assumed resolutions (FWHM) of 2\% and 3\% for the neutron and charged particle detectors, respectively \cite{Koh13}. The distributions for the half-lives shown in the figure demonstrate the approximate limits of the method. While the velocity difference distribution for a half-life of 1~ps is essentially centered around zero, the distribution for a half-life of 100~ps is centered at the edge of the target and distributions for longer half-lives are very similar.

    \begin{figure}
        \begin{center}
                \includegraphics[width=0.85\textwidth]{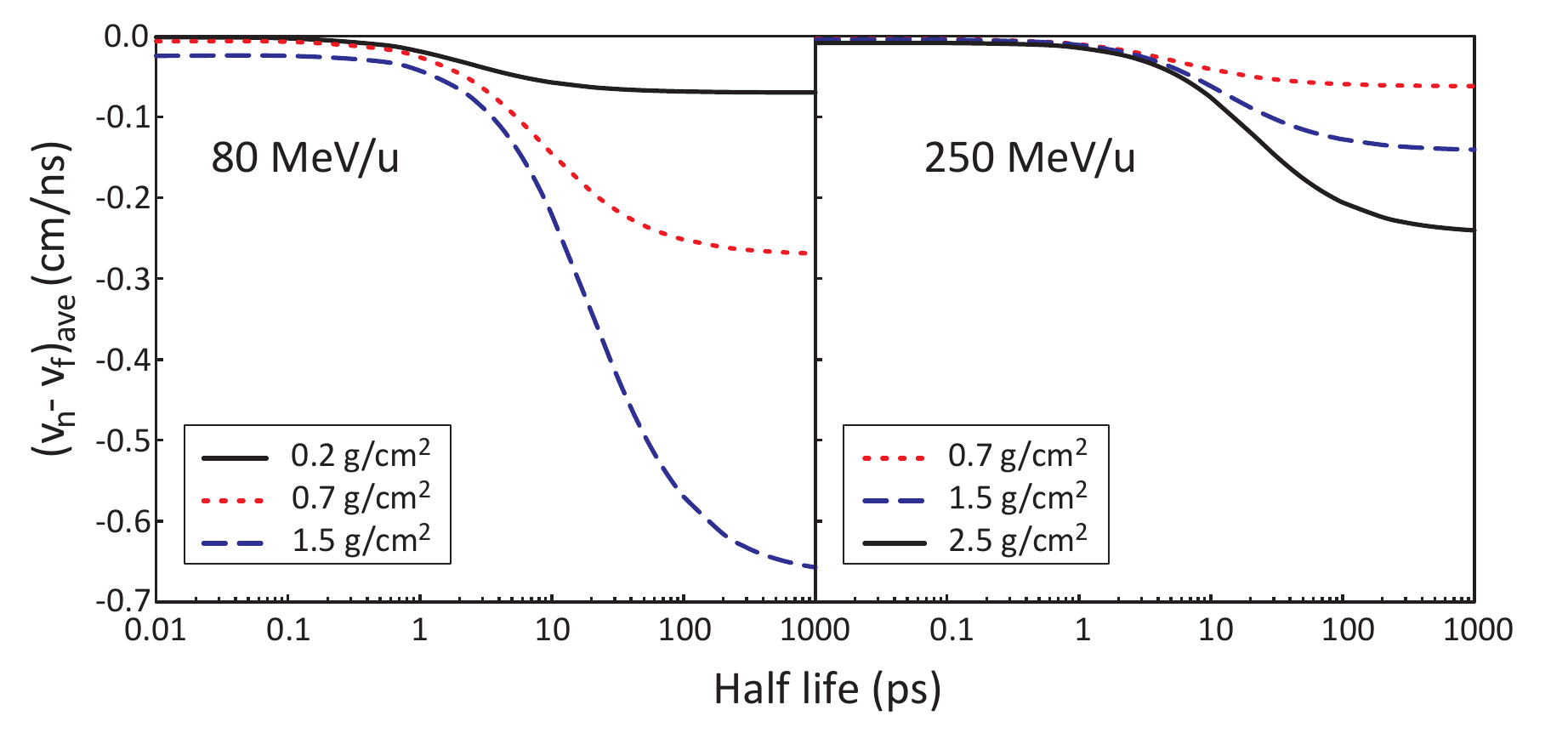}
        \end{center}
        \caption{Average value of the velocity difference distributions for the DiT technique as a function of half-life using the reaction $^{17}$C(-p)$^{16}$B$\rightarrow$$^{15}$B+n at 80 MeV/u (left) and 250 MeV/u (right) for three different target thicknesses.}
        \label{DiT:lifetime}
    \end{figure}

The general sensitivity of the method can be shown by plotting the average value of the folded velocity difference distributions as a function of half-lives. Figure \ref{DiT:lifetime} shows the dependence on the target thickness for incident beam energies of 80~MeV/u (a) and 250~MeV/u (b). The thicker targets are more sensitive because the velocity difference depends on the energy loss of the charged fragments in the target. For the same reason, the method is also more effective at lower energies for the same target thickness. 

    \begin{figure}
        \begin{center}
                \includegraphics[width=0.6\textwidth]{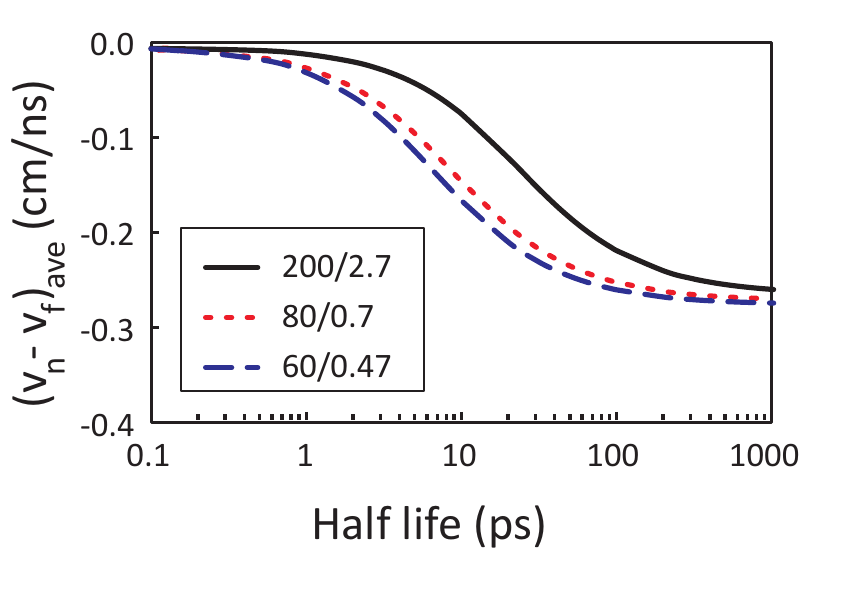}
        \end{center}
        \caption{Average value of the velocity difference distributions for the DiT technique as a function of half-life for the reaction $^{17}$C(-p)$^{16}$B$\rightarrow$$^{15}$B+n and three different combinations of beam energy and target thicknesses. The first number in the legend corresponds to the beam energy in MeV/u and the second number is the target thickness in g/cm$^2$.}
        \label{DiT:energy}
    \end{figure}

The choice of target thickness for a given energy is limited by the requirement that the remaining energy after the target has to be sufficiently large to allow for a clean identification of the fragment and a good energy measurement ($\sim$3\% in velocity, see above). Thus at higher energies, thicker targets can be used. Nevertheless, in order to push the methods to the smallest half-lives it is still advantageous to use lower beam energies as shown in figure \ref{DiT:energy}. The target thicknesses of 2700~mg/cm$^2$, 700~mg/cm$^2$, and 470~mg/cm$^2$ for beam energies of 200~MeV/u (black/solid line), 80~MeV/u (red/dotted line), and 60~MeV/u (blue/dashed line), respectively, were selected to reach approximately the same asymptotic value for very long half-lives. 

In addition to the target thickness and beam energy, the design of an experiment to search for neutron radioactivity has to consider other factors such as available beam intensity, reaction cross sections and detector resolution. The most important factor in order to extract a reasonable measurement of a half-life will most likely be statistics. At the present time, the beam intensities for very neutron-rich radioactive beams are still relatively small, although the new RIBF at RIKEN \cite{Yan07} has achieved some impressive increases. A recent experiment to study the two-neutron decay of $^{26}$O with the SAMURAI/NEBULA setup collected about a factor of 30 more statistics \cite{Kon12} than the published data from MSU/NSCL \cite{Koh13}. However, the experiment was performed at a higher incident beam energy (200~MeV/u) with a thicker target (2~g/cm$^2$) and it is not clear if the increased statistics will be sufficient to make up for the reduction in sensitivity in the $\sim$4~ps range (see figure \ref{DiT:energy}). If the beam energy at RIBF can be reduced without significant intensity losses SAMURAI/NEBULA is in the ideal position to improve on the lifetime measurement of $^{26}$O.

\section{Decay in Magnetic Field}

    \begin{figure}
        \begin{center}
                \includegraphics[width=0.9\textwidth]{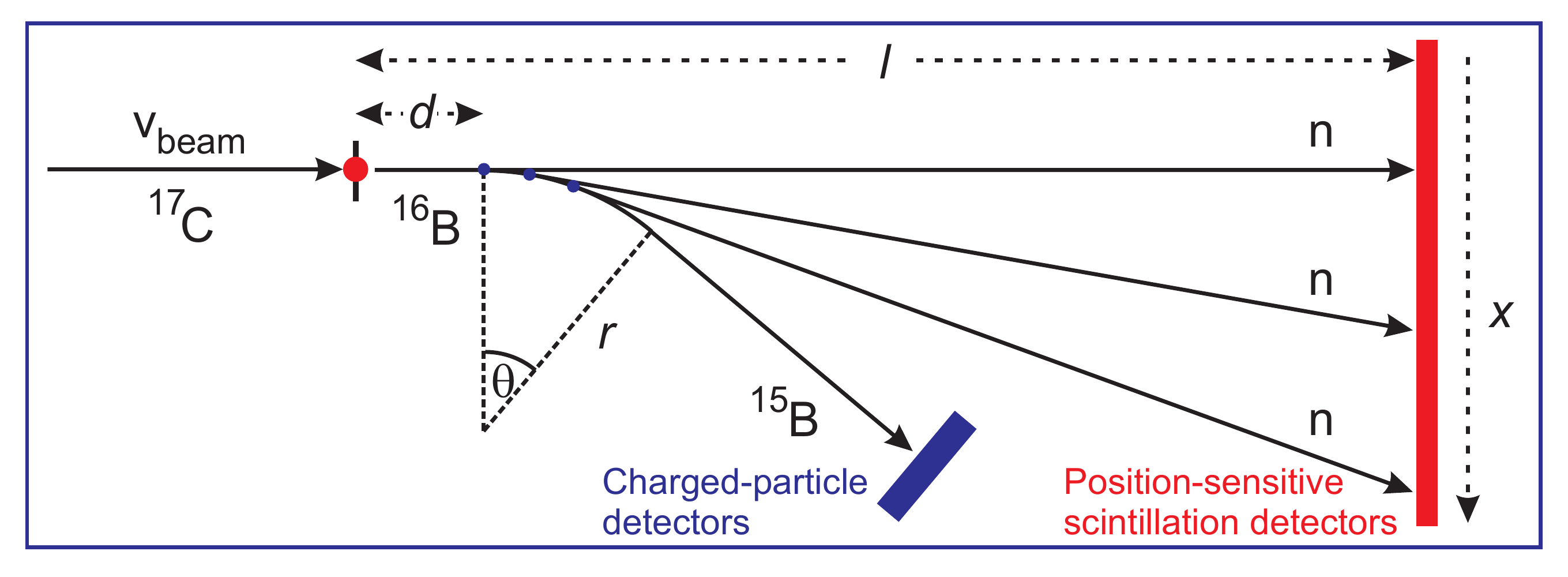}
        \end{center}
        \caption{Schematics of the DiMF technique. The incoming $^{17}$C beam produces $^{16}$B in the target (red dot) which enters a magnet field (bend radius $r$ and deflection angle $\theta$) after a drift distance $d$. $^{16}$B then can decay at different positions along the flight path (blue dots) into $^{15}$B and a neutron. The $^{15}$B fragments and neutrons are detected and identified in a set of charged-particle detectors and position sensitive scintillation detectors (at a distance $l$ from the target), respectively.}
        \label{DiMF:scheme}
    \end{figure}

An alternative method to measure finite lifetimes of neutron emitters proposed by Caesar {\it et al.} involves the deflection of the fragments in a magnetic field after the target and measuring the position distribution of the neutrons \cite{Cae12}. A schematic overview of such an experiment is shown in figure \ref{DiMF:scheme}. The reaction [$^{17}$C(-p)$^{16}$B] takes place in the target (red dot) and the subsequent decay ($^{16}$B$\rightarrow^{15}$B + n) can occur at later times in-flight (blue dots). A deflecting magnetic field with a bend radius $r$ and a deflecting angle $\theta$ bends the charged particles away from zero degrees after a possible drift distance $d$ into a set of charged-particle detectors. The neutrons are typically detected with an array of scintillation detectors \cite{Bla92,Bau05,Yon10,NEU12} located near zero degrees at a distance $l$ from the target. If the lifetime is sufficiently long for $^{16}$B to decay within the magnetic field, the horizontal distribution of the neutrons along the detector is directly related to the half-life.

    \begin{figure}
        \begin{center}
                \includegraphics[width=0.7\textwidth]{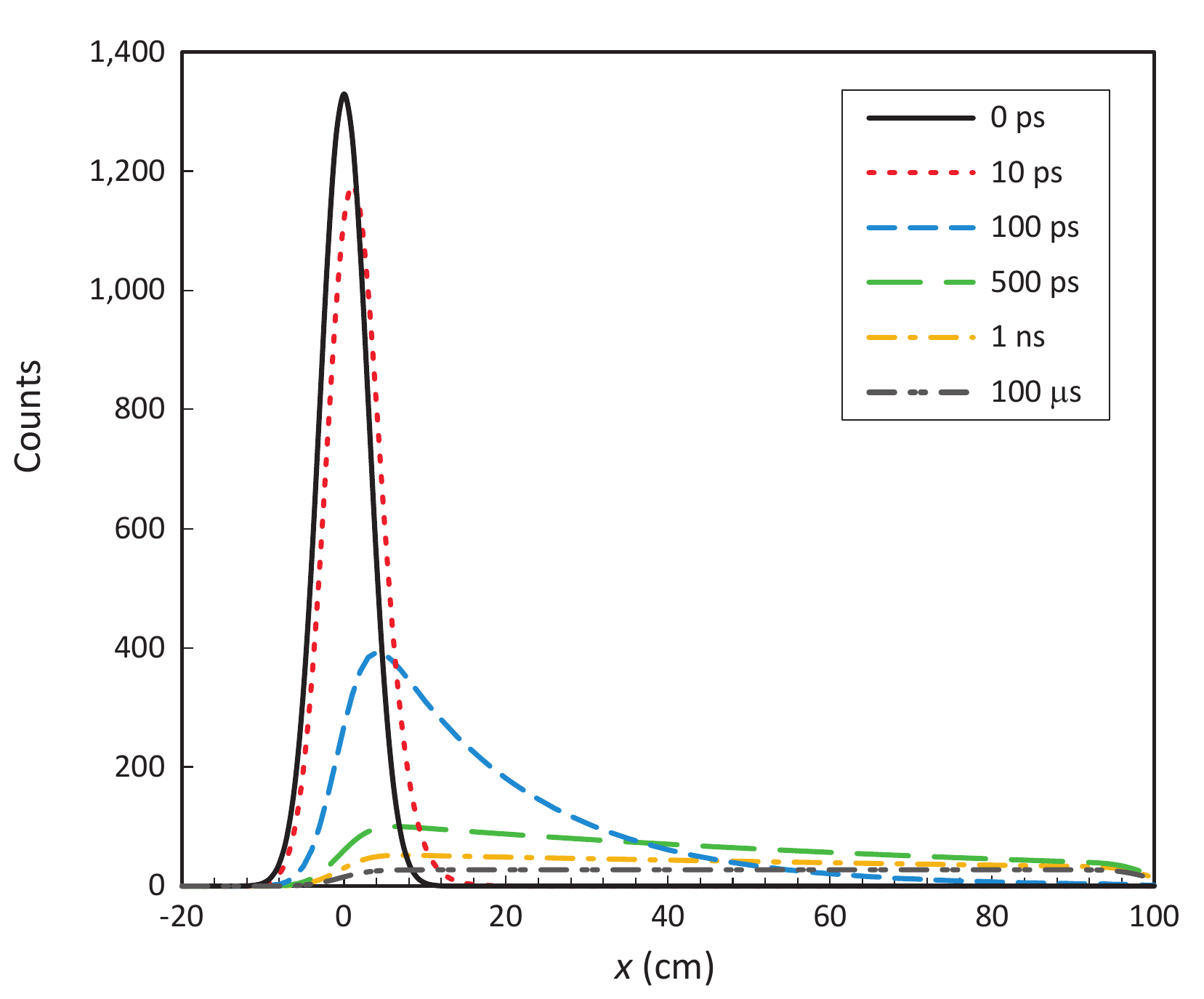}
        \end{center}
        \caption{Transverse position distributions for neutrons from the reaction $^{17}$C(-p)$^{16}$B$\rightarrow$$^{15}$B+n at 200 MeV/u for different half-lives. The drift distance, bend radius and neutron detector distances were 0~cm, 200~cm, and 1500~cm, respectively. }
        \label{DiMF:distributions}
    \end{figure}

Figure \ref{DiMF:distributions} shows these distributions for six different half-lives. The calculations were performed at a beam energy of 200~MeV/u and a bend radius of 2~m. The drift distance was 0~cm and 2-m long neutron detectors were placed at a distance of 15~m from the target and centered at the beam axis. The results were folded with a position resolution of 3~cm. 

Short lifetimes ($\lesssim$10~ps) result in a shift of the peak of the distributions, while for longer lifetimes, the exponential decay is directly proportional to an exponential position decay across the detector which can be expressed by a position decay constant: $\lambda_x$= $-$ln($\Delta$(Counts))/$\Delta$x). 
The optimum sensitivity for this arrangement is about 100~ps, where -- depending on the statistics of the experiment -- half-lives as short as 10~ps and as long as 1~ns might be measurable. 

\begin{figure}
        \begin{center}
                \includegraphics[width=0.51\textwidth]{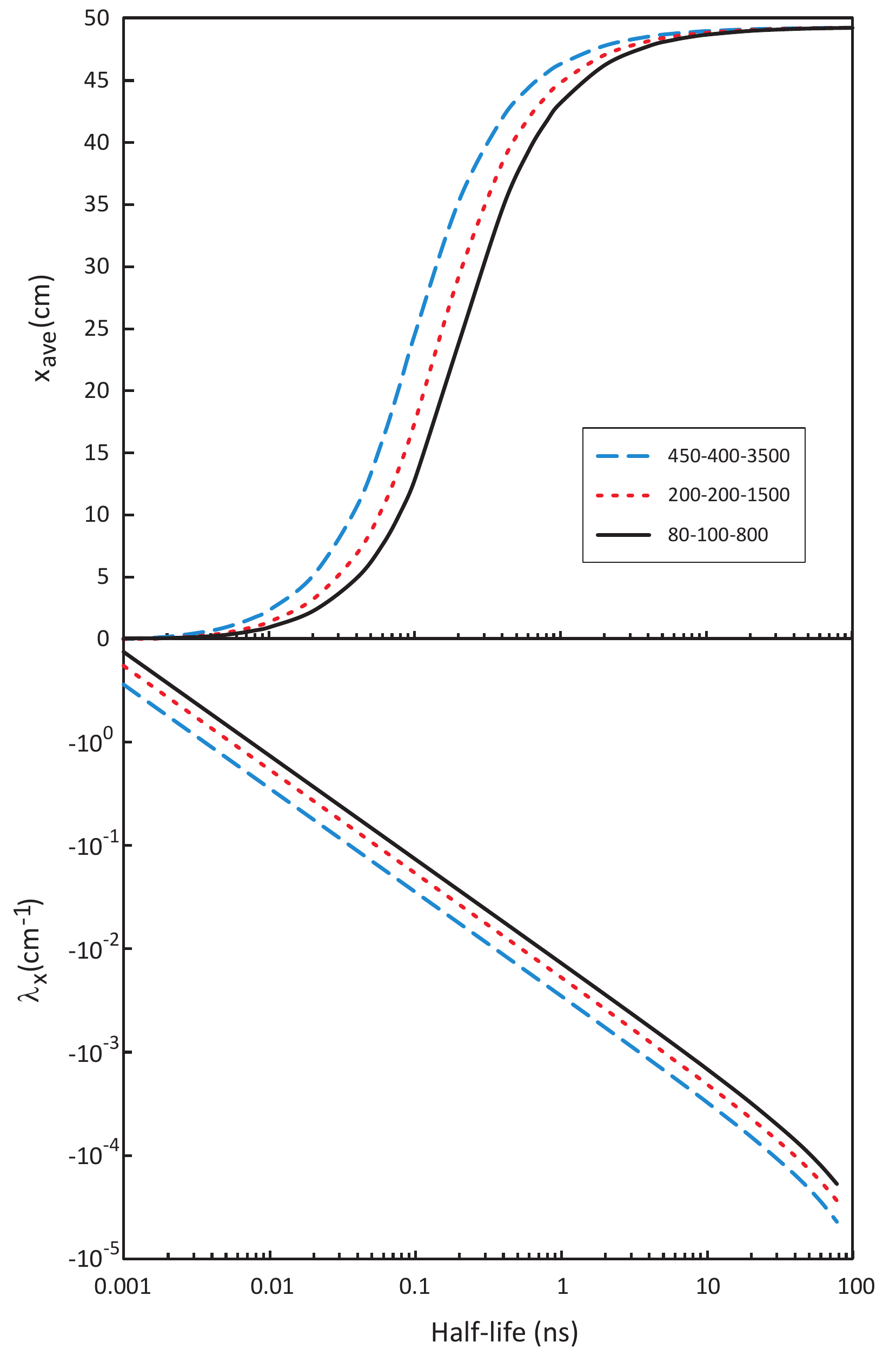}
        \end{center}
        \caption{Average position $x_{ave}$ (top) and position decay constant $\lambda_x$ (bottom) of the transverse position distributions for neutrons from the reaction $^{17}$C(-p)$^{16}$B$\rightarrow$$^{15}$B+n as a function of half-lives. The numbers in the legend box correspond to the incoming beam energy in MeV/u, bend radius in cm, and neutron detector distance in cm.}
        \label{DiMF:lifetime}
    \end{figure}

Similar to the DiT technique, in the actual experiment the half-life has to be extracted from a detailed fit to the whole distribution taking all experimental parameters into account. However, the sensitivity range of the DiMF method can be explored by analyzing the average value of the position distributions ($x_{ave}$) and the position decay constant $\lambda_x$ which are shown as a function of half-lives in the top and bottom panel of figure \ref{DiMF:lifetime}, respectively.

The parameters for the three lines shown were selected as approximate representations of the experimental setups Sweeper-MoNA at MSU/NSCL (black/solid) \cite{Bir05,Koh12}, SAMURAI-NEBULA at RIKEN/RIBF (red/dotted) \cite{SAM06}, and R$^3$B-NeuLAND at FAIR (blue/dashed) \cite{NEU12,Gas08}. The Sweeper-MoNA calculations were performed for an incoming beam energy of 80~MeV/u, a bend radius of 1~m, and a target to MoNA distance of 8~m. Calculations for the SAMURAI-NEBULA setup assumed a beam energy of 200~MeV, a bend radius of 2~m and a neutron time-of-flight distance of 15~m. The corresponding values for R$^3$B-NeuLAND were 450~MeV/u, 4~m and 35~m. It should be stressed that for all calculations the target was placed directly in front of the magnet with a drift distance of 0~m.

The most important parameter to reach sensitivities for short lifetimes is the beam energy. However, even at the highest beam energies it will be difficult to reach sensitivities below about 10~ps. The bend radius and the neutron time-of-flight distance are directly proportional, i.e. for a given beam energy a bend radius of 1~m and detector distance of 5~m is equivalent to a radius of 3~m and a distance of 15~m. 

While $x_{ave}$ is more sensitive to shorter lifetimes, $\lambda_x$ is more relevant for longer lifetimes. The sensitivity limits can be estimated by comparing the red curves in figure \ref{DiMF:lifetime} with the approximate parameters for the SAMURAI-NEBULA setup with the distributions of figure \ref{DiMF:distributions} which were calculated with the same parameters. The lower limit for extracting a half-life of $\sim$10~ps is determined from the lower limit of measuring the shift of $x_{ave}$ at about 2~cm, while the upper limit of $\sim$1~ns is due the limit of measuring $\lambda_x$ at about $-$0.01~cm$^{-1}$.

    \begin{figure}
        \begin{center}
                \includegraphics[width=0.95\textwidth]{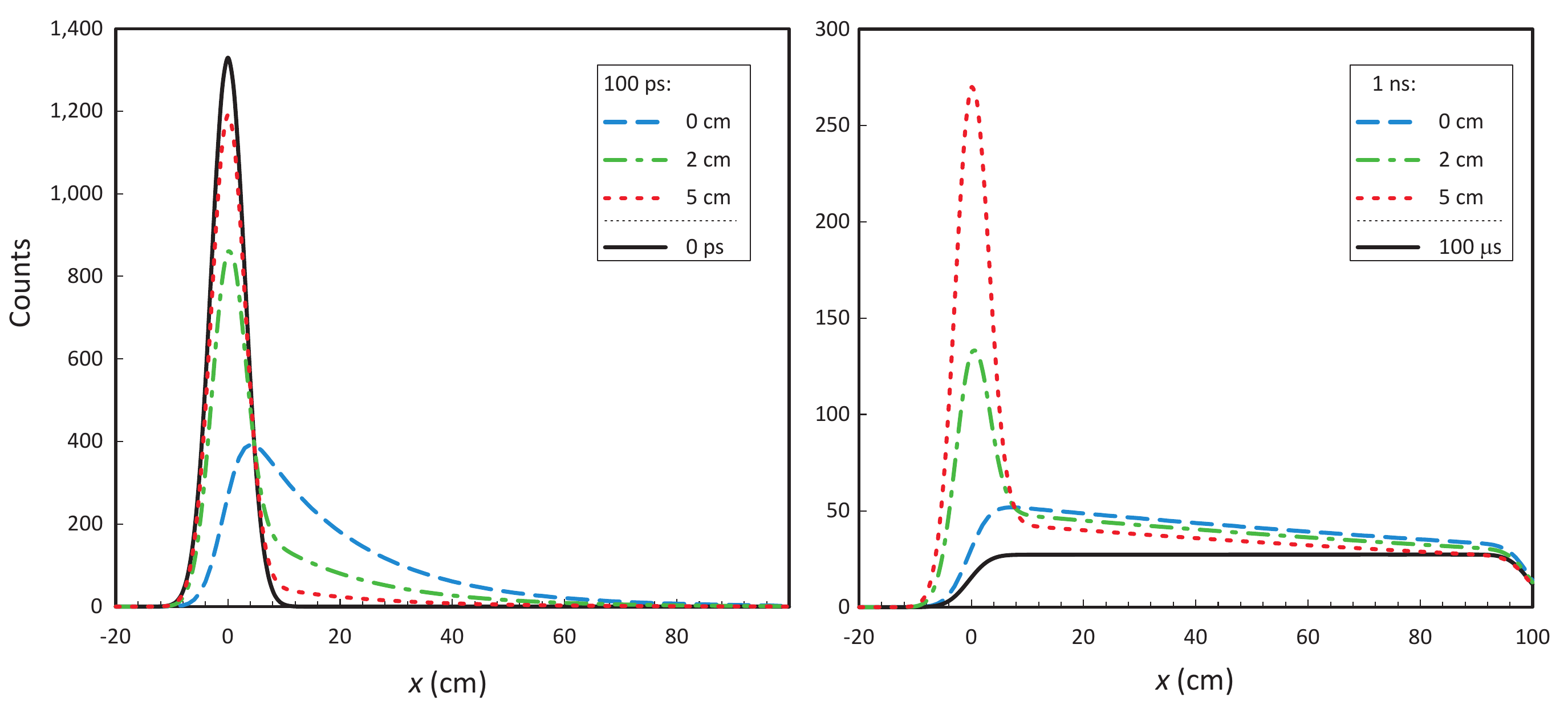}
        \end{center}
        \caption{Transverse position distributions for neutrons from the reaction $^{17}$C(-p)$^{16}$B$\rightarrow$$^{15}$B+n at 200 MeV/u and half-lives of 100~ps (left) and 1~ns (right). Distributions for drift distances of 0~cm (blue/dashed line) 2~cm (green/dot-dashed line) and 5~cm (red/dotted line) are shown.  For reference, the distributions for half-lives of 0~ps and 100~$\mu$s for 0~cm drift distance (black/solid line) are included in the left and right panel, respectively. The calculations were performed for the same number of reactions. Note the difference in the vertical scales.}
        \label{DiMF:drift}
    \end{figure}

The importance of placing the target directly in front of the deflecting magnet is demonstrated in figure \ref{DiMF:drift}. Especially for the shorter lifetimes, it will be critical to locate the target as close as possible to the magnet. As shown in the left panel of the figure, for a half-life of 100~ps a drift distance as small as 2~cm already moves the peak of the distribution to the center with only a tail extending to larger distances. For longer lifetimes the drift distance is not as critical. Although a certain fraction of the events will accumulate in a peak at the center, the position decay constant of the distribution remains the same and the overall intensity of the tail is not significantly reduced (see the right panel of the figure). 

\section{Conclusions}

Two methods to search for neutron radioactivity were discussed. In the first method the velocity difference of the neutrons and the charged fragments is measured. This method is sensitive to decays that occur within the target (DiT) and depends on the energy loss of the charged fragment in the target. The second method is sensitive to decays in-flight after the target by relying on the deflection of the charged fragment in a magnetic field (DiMF) which will then translate into the horizontal distribution of the neutrons. In order to be sensitive to the shortest lifetimes, the DiT method is better at low beam energies while the DiMF method is more sensitive at high beam energies. Within the range of presently available beam energies and experimental setups the DiT method is more sensitive to half-lives in the 1--100~ps range, while the DiMF method is sensitive to half-lives between $\sim$10~ps and $\sim$1~ns.

\section*{Acknowledgments}

This work was supported by the National Science Foundation under grant PHY-11-02511. We would like to thank P. DeYoung, J. E. Finck, R. Haring-Kaye, and S. Stephenson for valuable comments and careful reading of the manuscript.

    \bibliographystyle{nphys}

\end{document}